\documentclass[aps,prb,twocolumn,showpacs,superscriptaddress]{revtex4-1}
\usepackage{graphicx}
\usepackage{xcolor}   
\usepackage{amsmath}
\usepackage{epsfig}

\newcommand{\yy}{y}
\newcommand{\im}{a}
\newcommand{\jm}{b}

\newcommand{\Ge}{G_e}
\renewcommand{\kappa}{\sigma}
\newcommand{\tk}{\tilde{K}}
\newcommand{\vn}{\vec{\textbf{n}}}
\newcommand{\vk}{\vec{\textbf{k}}}
\begin{document} 
\title{Quantum thermal transport in stanene}
\author{Hangbo Zhou}
\affiliation{Institute of High Performance Computing, A*STAR, 138632 Singapore}
\author{Yongqing Cai}
\affiliation{Institute of High Performance Computing, A*STAR, 138632 Singapore}
\author{Gang Zhang}
\email[]{zhangg@ihpc.a-star.edu.sg}
\affiliation{Institute of High Performance Computing, A*STAR, 138632 Singapore}
\author{Yong-Wei Zhang}
\affiliation{Institute of High Performance Computing, A*STAR, 138632 Singapore}
\date{\today}

\begin{abstract}
By way of the non-equilibrium Green's function simulations and analytical expressions, the quantum thermal conductance of stanene is studied. We find that, due to the existence of Dirac fermion in stanene, the ratio of electron thermal conductance and electric conductance becomes a chemical-potential-dependent quantity, violating the Wiedemann-Franz law. This finding is applicable to any two-dimensional (2D) materials that possess massless Dirac fermions. In strong contrast to the negligible electronic contribution in graphene, surprisingly, the electrons and phonons in stanene carry a comparable heat current. The unusual behaviours in stanene widen our knowledge of quantum thermal transport in 2D materials.
\end{abstract}
\pacs{44.10.+i, 65.80.-g, 63.22.-m, 73.43.-f}
\maketitle

\section{Introduction}

Stanene, a monolayer of tin atoms that was fabricated recently \cite{a1}, is a promising material for the realization of novel quantum devices due to its striking electronic properties \cite{Xu2014,st1,st2,st4,st5,Xu2013,st3}. For instance, the strong spin-orbital coupling (SOC) in stanene is able to open a large enough band gap, which may be suitable for application as room-temperature quantum spin-hall (QSH) insulators \cite{Xu2013,st3, st6,st7,st8,st9}. Recently it was also shown that it could be a promising thermoelectric material due to its non-dissipative conduction edges \cite{st1}.

Besides electrical conduction, thermal conduction is another important form of energy and information transport \cite{RMP}. In general, thermal transport is carried by either electrons or phonons. Many nanomaterials, such as graphene and carbon nanotubes \cite{Balandin2008,Balandin2011,Munoz2010,Balandin,ShiL,AIPadvance, Saito2007,Zhang2005,Liu2012}, have significantly higher phonon thermal conductance than their electronic counterpart at room temperature. However, in other materials, such as metals, the electron contribution dominates the thermal transport. The electron thermal conductance is governed by the Wiedemann-Franz (WF) law, which imposes a universal relation between electron thermal conductance $\kappa_e$ and the electronic conductance $\Ge$ as, $\kappa_e=L_0\Ge T$, where $T$ is temperature and $L_0=\frac{\pi^2}{3}(\frac{k_B}{e})^2$ is a fundamental constant derived from Sommerfeld theory based on low temperature expansion\cite{Ashcroft2011}.  

The electronic properties of 2D materials have been studied extensively \cite{Neto2009,Peres2010,Munoz2012,Neto2009,Crossno2016}. Many interesting phenomena of 2D materials, such as massless Dirac fermion \cite{Peres2010, Crossno2016}, spin-orbital coupling \cite{Neto2009}, electron-phonon interaction \cite{Munoz2012} and their effects on the electronic transport properties have been examined. Compared with graphene, the transport properties, in particular the thermal transport of stanene, have not been extensively investigated. Many interesting and important properties remain unexplored. For instance, whether electrons or phonons dominate the thermal transport is yet unknown. The applicability of WF law to stanene, as well as other 2D QSH insulators, has not been examined. Clearly, answers to these questions are not only of scientific interests in understanding the transport mechanisms in 2D Dirac fermions systems, but also of technological significance to the applications of stanene-based quantum devices.  

In this article we study quantum thermal transport of stanene in ballistic transport regime by using the non-equilibrium Green's function (NEGF) approach, which has been widely used in the study of transport properties of graphene. We find that electron thermal conductance of massless Dirac fermions is proportional to its electronic conductance in a large range of temperature. However, the proportionality constant remarkably depends on the chemical potential, which signifies the breakdown of the conventional WF law. In addition, we also derived an analytical formula for the ratio of these two conductances, which is applicable to Dirac fermions materials in general. It is found the electron thermal conductance is substantially important in stanene in comparison with phonon counterpart, and it can even become dominant at room temperature when it is gated.

\section{Results and discussion}
\vspace{10pt}
\label{Phonon thermal conductance}
\subsection{Phonon thermal conductance} 
\begin{figure}
\includegraphics[width=\linewidth]{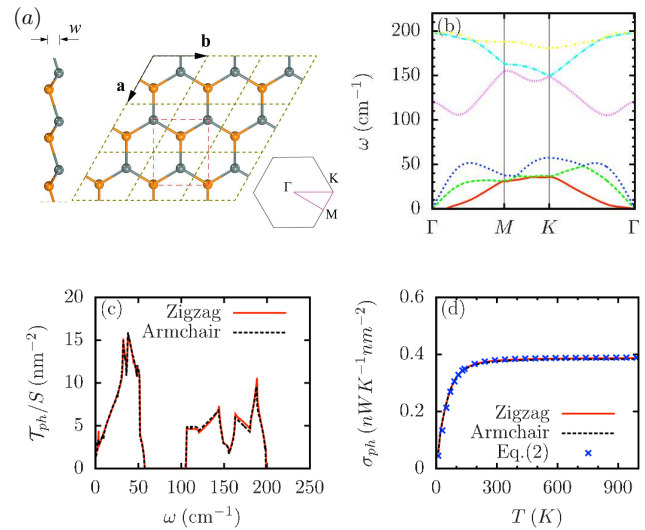}
\caption{\label{fig:phonon}(a) Crystal structure of stanene and its high-symmetry points in reciprocal points. Atoms in different planes are labelled with different colors. (b) The phonon dispersion relation of stanene. (c) Plot of transmission function per cross-sectional area against phonon frequency. (d) Plot of thermal conductance versus temperature along both zigzag and armchair directions. We estimated $\bar{\mathcal{T}}/S=6nm^{-2}$ in Eq.~(\ref{eq:kph}).}
\end{figure}

In the ballistic transport regime, the phonon thermal conductance is given by Landauer formula
\begin{equation}
\kappa_{ph}=k_B\int_0^{\omega_m}\frac{d\omega}{2\pi S} \mathcal{T}_{ph}[\omega]\frac{(\beta\hbar\omega)^2 e^{\beta\hbar\omega}}{(e^{\beta\hbar\omega}-1)^2},
\end{equation} 
where $\beta=1/(k_BT)$ denotes the inverse temperature. $\omega_m$ is the maximum possible frequency of phonon modes and $\mathcal{T}_{ph}$ is the phonon transmission function through cross area $S$. For convenience, we choose $S=a\times W$, where $a$ is the lattice constant of a unit cell in the transverse direction, and $W$ is the thickness of stanene. We evaluate $\mathcal{T}_{ph}[\omega]$  using NEGF formalism based on the interatomic force constants\cite{Zhang2007,Wang2008}. The details are presented in Appendix. The force constants are obtained by first-principles calculations using Quantum Espresso \cite{QE} within the scheme of density functional perturbation theory. The norm-conserving pseudopotential and local density approximation of Ceperley-Alder are adopted together with a cutoff energy of 68 Ry. The relaxed lattice constant of stanene is 4.523\AA$\,$and the buckling height of stanene layer is 0.822\AA, as illustrated in Fig.~\ref{fig:phonon}(a). The thickenss of stanene is $W=3.954$\AA, by adding the diameter of a tin atom to the buckling height. We sample the Brillouin zone with a $18\times 18\times 1$ Monkhorst-Pack $k$-grid. The atomic coordinates are fully relaxed until the forces become smaller than $1\times 10^{-6}$eV/\AA. An $8\times 8\times 1$ $q$-mesh is used to calculate the dynamic matrix for inverse Fourier transformation to obtain the force constants in real space. According to the interatomic force constants, we evaluate the phonon dispersion relation Fig.~\ref{fig:phonon}(b), phonon transmission function [Fig.~\ref{fig:phonon}(c)] and phonon thermal conductance Fig.~\ref{fig:phonon}(d). As indicated in both plots of dispersion relation and transmission function, the maximum phonon frequency is around $200\mbox{cm}^{-1}$, which is much smaller than other 2D materials such as graphene ($1500\mbox{cm}^{-1}$) \cite{Balandin,ShiL}, MoS$_2$ ($500\mbox{cm}^{-1}$ \cite{a2,a3} and phosphorene ($480\mbox{cm}^{-1}$) \cite{refa4,a6}. A large gap exists between the acoustic modes and optical modes. Stanene is found to be highly isotropic as the transmission in the zigzag and armchair directions differs little. Therefore, in the following discussions, we focus only on the transport properties in the zigzag direction. From the thermal conductance plot, we find that $\kappa_{ph}$ increases with the increase of temperature and saturates at around 200K. The saturated thermal conductance is around  0.39$n\mbox{WK}^{-1}nm^{-2}$, which is much smaller than that of graphene (4.1$n\mbox{WK}^{-1}nm^{-2}$) \cite{a5}.

In order to give an explicit form of temperature dependence of phonon thermal conductance, we approximate the transmission function by its average value $\bar{\mathcal{T}}$. Physically, $\bar{\mathcal{T}}$ can be estimated from dispersion relationship $\omega(k_\parallel,k_\perp)$, where $k_\parallel$ is the wavevector in the transverse direction and $k_\perp$ is that in the transport direction. For each branch we can project the dispersion relation onto the plane spanned by $k_\parallel$ and $\omega$, and the total area of the resulting image is $\bar{\mathcal{T}}\omega_m$. This approximation will break down in the low temperature limit, and it converges to the exact value in high temperature limit.
Using this approximation, the phonon thermal conductance is found to be
\begin{equation}
\label{eq:kph}
\kappa_{ph}=\frac{k_B\bar{\mathcal{T}}\omega_m}{2\pi Sx}\Big[\frac{\pi^2}{3}-x^2\mbox{Li}_0(e^{-x})-2x\mbox{Li}_1(e^{-x})-2\mbox{Li}_2(e^{-x})\Big],
\end{equation}
where $x=\beta\hbar\omega_m$ and $\mbox{Li}_n(z)=\sum_{k=1}^\infty z^k/k^n$ are the polylogarithm functions \cite{Lee1995}. The maximum frequency $\omega_m$ can be estimated from the largest force constant $K_m$ and the corresponding atomic mass $m$, via $\omega_m=\sqrt{2K_m/m}$. From Fig.~\ref{fig:phonon}(d) we see that the curve from the estimation formula matches well with the NEGF curve. In the high temperature limit, the thermal conductance approaches 
\begin{equation}
\kappa_{ph}=k_B\bar{\mathcal{T}}\omega_m/(2\pi S).
\end{equation}

\vspace{10pt}
\label{Electronic properties}
\subsection{Electronic transport} 
The electronic structure of stanene can be modelled by tight-binding Hamiltonian as established in Ref.~\cite{Liu2011}. In this work, we consider the transport properties using the electron structure with spin-orbital coupling (w/SOC) and without spin-orbital coupling (w/o SOC). The tight-binding Hamiltonian can be written as
\begin{equation}
H=H_0\otimes I_2+H_{SO},
\end{equation}
where $H_0$ is the Hamiltonian without SOC,  $I_2$ is a $2\times 2$ identity matrix due to spin degeneracy and $H_{SO}$ is the SOC Hamiltonian. $H_0$ can be written as the sum of an on-site term and a hopping term
\begin{equation}
H_0=\sum_{i,\alpha}E_{\alpha}c^\dagger_{i,\alpha}c_{i,\alpha}+\sum_{\left<ij\right>,\alpha,\beta}t_{\left<ij\right>,\alpha\beta}c^\dagger_{i,\alpha}c_{j,\beta},
\end{equation}
where $c_{i,\alpha}(c^\dagger_{i,\alpha})$ is an electron annihilation (creation) operator of site $i$ with orbital states $\alpha\in\{s,p_x,p_y,p_z\}$. The onsite energy for each orbital is $E_{\alpha}=\{\varepsilon_s,\varepsilon_p,\varepsilon_p,\varepsilon_p\}$. The factor $t_{\left<ij\right>,\alpha\beta}$ is the hopping constant between nearest neighbours $i$ and $j$. If they are in the direction of a norm vector $r_{\left<ij\right>}=\{r_x,r_y,r_z\}$,  this hopping constant is given by the Slater-Koster formulas \cite{Slater1954}: $t_{ss}=V_{ss\sigma}$, $t_{sp_\im}=r_\im V_{sp\sigma}$, $t_{p_\im p_\im}=r_\im^2V_{pp\sigma}+(1-r_\im^2)V_{pp\pi}$ and $t_{p_\im p_\jm}=r_\im r_\jm(V_{pp\sigma}-V_{pp\pi}), \im \neq \jm$, where $\im,\jm\in\{x,y,z\}$ are polarization indices.

The SOC Hamiltonian can be written as a product of angular momentum and spin operator \cite{Konschuh2010}. Explicitly, it is
\begin{equation}
H_{SO}=\xi/2\sum_{i,\alpha,\beta,\sigma,\sigma'}-i\epsilon_{\alpha\beta\gamma}\,\, c^\dagger_{i,\alpha,\sigma}(\sigma_\gamma)_{\sigma\sigma'} c_{i,\beta,\sigma'},
\end{equation}
where $\sigma,\sigma'$ are the spin indices representing up or down. The orbital indices $\alpha,\beta,\gamma\in\{p_x,p_y,p_z\}$ now exclude $s$ ($s$ orbitals are still spin-degenerate) and $\epsilon_{\alpha\beta\gamma}$ is the Levi-Civita symbol. $\vec{\sigma}=\{\sigma_x,\sigma_y,\sigma_z\}$ is the Pauli matrix. 

\begin{figure}
	\includegraphics[width=\linewidth]{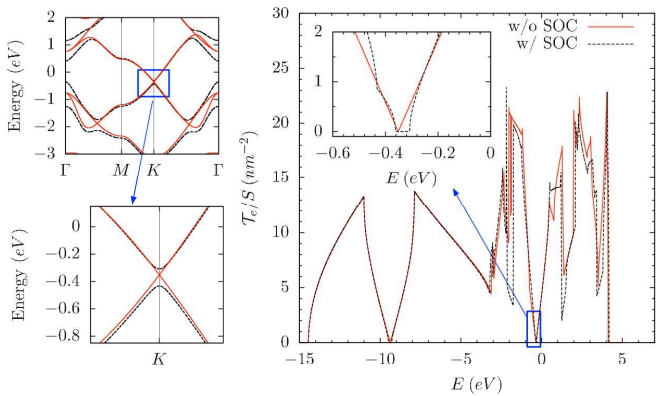}
	\caption{\label{fig:electron}Band structure (left) and transmission function per cross-sectional area (right) of stanene. The parameters we used are $V_{ss\sigma}=-2.62eV$, $V_{sp\sigma}=2.65eV$, $V_{pp\sigma}=1.49eV$, $V_{pp\pi}=-0.79eV$, $\varepsilon_s-\varepsilon_p=-6.23eV$ \cite{Pedersen2010}, and $\xi=0.8eV$ \cite{Chadi1977}. }
\end{figure}

This tight-binding Hamiltonian model has been used in combination with first-principle calculations \cite{Liu2011} to study electronic properties. Figure~\ref{fig:electron} (left) shows the bandstructure of stanene predicted by this Hamiltonian using the parameters given in Ref.~\cite{Liu2011}. It produces a reasonable bandstructure in comparison with density-functional-theory calculations \cite{Xu2013}. When SOC is not considered, a Dirac cone is predicted at $K$ point. Once SOC is introduced, a band-gap is opened at the Dirac cone.

The electronic transport properties are determined by the Onsager transport coefficients
\begin{equation}
\label{eq:Landauer}
L_n=\frac{1}{\hbar}\int\frac{dE}{2\pi S}\mathcal{T}_e[E]\frac{\beta(E-\mu)^n e^{\beta(E-\mu)}}{[e^{\beta(E-\mu)}+1]^2},
\end{equation}
where $\mathcal{T}_e[E]$ is the electron transmission function including the spin factor, and $\mu$ is the chemical potential. Based on the transport coefficients, the electronic conductance $\Ge$ and electron thermal conductance $\kappa_e$ can be evaluated via $\Ge=e^2L_0$ and $\kappa_e=(L_0L_2-L_1^2)/(L_0T)$. The energy-parametrized transmission function is evaluated by integrating over all allowed modes \cite{Zhang2007}
\begin{equation}
\label{eq:TE}
\mathcal{T}_e(E)=a\int_{k_{\|}\in \mbox{\footnotesize{BZ}}}\frac{dk_{\|}}{2\pi}\Xi(E,k_{\|}),
\end{equation}
where $k_{\|}$ is the wave vector in the transverse direction of transport, $a$ is the lattice constant in the transverse direction of transport and $\Xi(E,k_{\|})$ is the transmission function of a mode with energy $E$ and transverse wave-vector $k_{\|}$. In a perfect lattice, $\Xi(E,k_{\|})$ is an integer that counts the number of allowed modes. 

\vspace{10pt}
\label{Wiedemann-Franz law}
\subsection{Modified Wiedemann-Franz law}  
\begin{figure}
	\includegraphics[width=\linewidth]{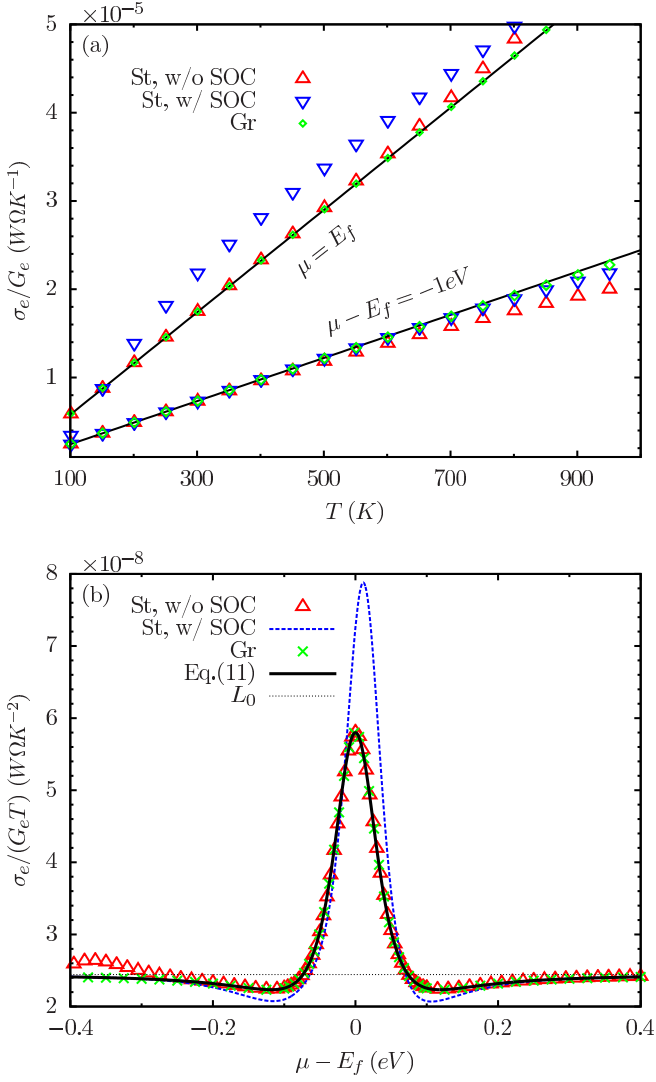}
	\caption{\label{fig:wf}The conductance ratio in stanene. (a) Temperature profile of conductance ratio. The black solid lines represents slope of $5.79\times10^{-8}W\Omega K^{-2}$ and $2.44\times10^{-8}W\Omega K^{-2}$ respectively. (b) Plots of conductance ratio against the chemical potential at $T=300K$. The bottom solid black line is the traditional Lorenz number $L_0=2.44\times10^{-8}W\Omega K^{-2}$.  In all above plots, the tight-binding parameters for stanene are the same as those in Fig.~\ref{fig:electron}. The parameters of tight-binding graphene is taken from Ref.~\cite{Saito1992}, as $V_{ss\sigma}=-6.77eV$, $V_{sp\sigma}=5.58eV$, $V_{pp\sigma}=5.04eV$, $V_{pp\pi}=-3.03eV$, and $\varepsilon_s-\varepsilon_p=-8.87eV$. SOC in graphene is ignored.}
\end{figure}

We first study the conductance ratio of stanene $\kappa_e/\Ge$ against temperature, as shown in Fig.~\ref{fig:wf}(a). It is interesting to see that the conductance ratio increases with temperature linearly, in the range up to 900K. However, the proportionality is not a fundamental constant. In fact, it depends significantly on the chemical potential. When the chemical potential is shifted away from the Dirac point, the proportionality is close to $L_0=2.44\times 10^{-8}W\Omega K^{-2}$, which recovers the conventional WF law. However, when $\mu$ is at the Fermi-level $\mu= E_f=-0.351eV$, the linear relationship still holds, but the proportionality becomes much larger than the Lorenz constant. The same phenomenon is also found in graphene as well (green symbols). Importantly, we observe that the conductance ratio at Dirac point for stanene (w/o SOC) and graphene matches each other, implying that there exists an intrinsic law of conductance ratio for Dirac fermions. 

In order to reveal the chemical potential dependence, we explicitly plot $\kappa_e/(\Ge T)$ against $\mu-E_f$ in Fig.~\ref{fig:wf}(b) at room temperature.  The Lorenz number $L_0\approx 2.44\times 10^{-8}W\Omega K^{-2}$, is shown as a reference in the bottom. It is obvious that for both graphene and stanene, the value of $\kappa_e/(\Ge T)$ approaches the Lorenz constant, when $|\mu-E_f|$ is large. However, when $\mu=E_f$ , a much larger value of $\kappa_e/(\Ge T)$ is observed. 

In order to understand this novel feature in 2D QSH materials, we derive an analytical formula of conductance ratio using NEGF by focusing on the Dirac point. An important feature of Dirac Fermion is that its energy becomes proportional to the wave vector $\vec{k}$ from all directions $E_{\vec{k}}=\hbar v_F |\vec{k}|$, where $v_F$ is the Fermi velocity. Here we have shifted the center to the $K$ point and set $E_F=0$ for notation convenience. For each given energy $E$, the $k$-space forms a circle of radius $|E|/(v_F\hbar)$. As a result  for each $k_{\|}$ under condition $|k_{\|}|<|E|/(v_F\hbar)$, their exist two modes with different $k_\perp$. By considering the spin-degeneracy, there exist 4 modes in total of given $k_{\|}$ and $E$. It suggests $\Xi[E,k_{\|}]=4$ if and only if $|k_{\|}|<|E|/(v_F\hbar)$. Then according to Eq.~(\ref{eq:TE}), the transmission function near the Dirac point becomes proportional to energy
 \begin{equation}
\label{eq:TeE}
\mathcal{T}_e(E)=4a|E-E_f|/(\pi\hbar v_F).
\end{equation}
This proportionality feature is confirmed from the NEGF numerical results as shown in Fig.~\ref{fig:wf}(b). By plugging Eq.~(\ref{eq:TeE}) into Eq.~(\ref{eq:Landauer}), we can obtain
\begin{equation}
L_n=\frac{2C_n}{\pi^2\hbar^2 v_F W}(k_B T)^{n+1}\equiv\eta C_n(k_B T)^{n+1},
\end{equation}
where $\eta=2/(\pi^2\hbar^2 v_F W)$ is a shorthand notation. Importantly, $C_n$ is a function of $\mu$ as $C_n=\int_{-\infty}^{+\infty}dx\frac{|x+\beta(\mu-E_f)|x^ne^x}{(e^x+1)^2}$ with $x=\beta(E-\mu)$. By setting $y=\beta(\mu-E_f)$, one can find
\begin{eqnarray}
&C_0&=\yy-2\mbox{Li}_1(-e^{-\yy}),\nonumber\\
&C_1&=\pi^2/3+2\yy\mbox{Li}_1(-e^{-\yy})+4\mbox{Li}_2(-e^{-\yy}),\nonumber\\
&C_2&=\yy\pi^2/3-2\yy^2\mbox{Li}_1(-e^{-\yy})-8\yy\mbox{Li}_2(-e^{-\yy})-12\mbox{Li}_3(-e^{-\yy}).\nonumber
\end{eqnarray}
Hence the conductance ratio becomes
\begin{equation}
\label{eq:wfLd}
\frac{\kappa_e}{\Ge}=\frac{C_2C_0-C_1^2}{C_0^2}\left(\frac{k_B}{e}\right)^2T\equiv L_D T.
\end{equation}
When $y\gg 1$, or $\mu-E_f\gg k_BT$, the polylogarithm functions decay to 0 and $(C_2C_0-C_1^2)/C_0^2\rightarrow \pi^2/3$, hence,  the convention WF law is recovered analytically. This is because this procedure is equivalent to the Sommerfeld low temperature expansion. However, at Dirac point $\mu=E_f$, or $y=0$, the low temperature expansion breaks down and the polylogarithm functions become important. The conductance ratio turns out to be 
\begin{equation}
L_D\Big|_{\mu=E_f}=\frac{6\mbox{Li}_3(-1)}{\mbox{Li}_1(-1)}\left(\frac{k_B}{e}\right)^2\approx 5.79\times10^{-8}W\Omega K^{-2}.
\end{equation}
This constant is about two times larger than the Lorenz number and it is a general law for Dirac fermions. It also predicts that the width of the peak presented in Fig.~\ref{fig:wf}(b) is determined only by temperature.

We find that the analytical result of the proportionality $L_D$ in Eq.~(\ref{eq:wfLd}) matches well to the NEGF numerical calculations, in both temperature profile [Fig.~\ref{fig:wf}(a)] and chemical profile [Fig.~\ref{fig:wf}(b)]. Importantly, they matches for both stanene and graphene. The conductance ratio of stanene with SOC deviates from the modified WF law due to the lack of Dirac point, but it gives a reasonable result in comparison with the traditional one. In all the cases, the agreement becomes worse at high temperature end since the high energy electrons start to play a role.

\begin{figure}
	\includegraphics[width=\linewidth]{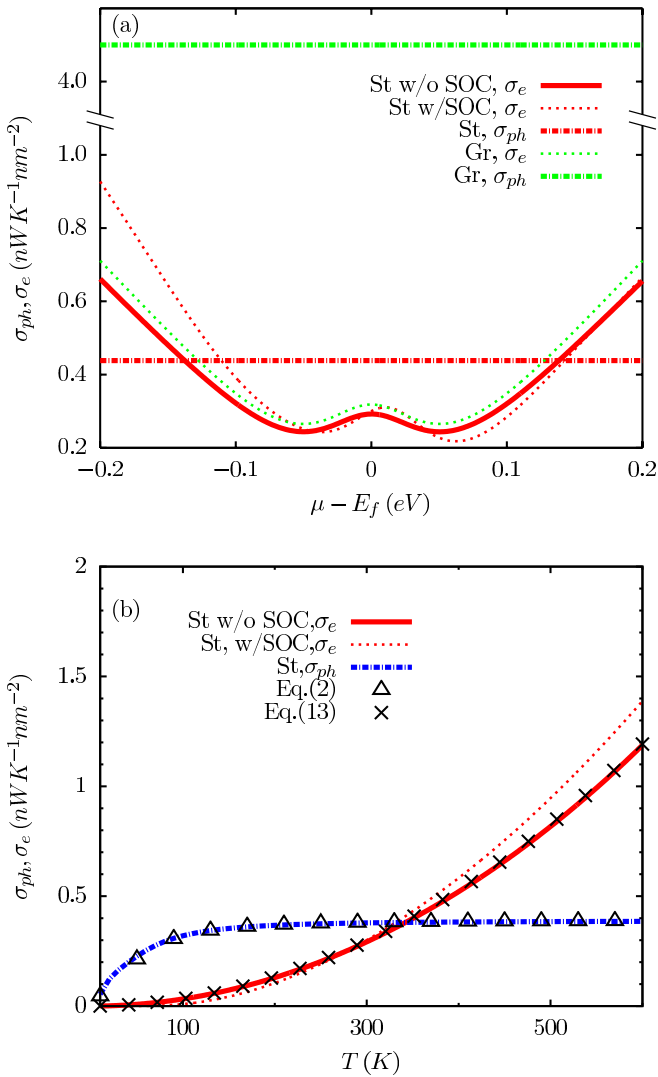}
	\caption{\label{fig:thermal}(a) Plot of electronic and phonon thermal conductance of stanene (St, red lines) and graphene (Gr, green lines) against chemical potential.  The temperature is set to be $T=300K$. (b) Plots of electron and phonon thermal conductance with respect to temperature at $\mu=E_f$. We use $v_F=4\times10^5$m/s obtained from dispersion relation and $\bar{\mathcal{T}}=0.5$ from transmission function. The thickness of graphene sheet is 1.7\AA and the SOC of graphene is negelected.}
\end{figure}

\vspace{10pt}
\label{Comparison}
\subsection{Comparison between phonon and electron thermal conductance.} Next, we make a comparison between phonon and electron thermal conductance, as shown in Fig.~\ref{fig:thermal}(a). We assume that $\kappa_{ph}$ is not sensitive to the chemical potential. For graphene (green lines), we find that $\kappa_{ph}$ is much larger than $\kappa_e$. When graphene is not gated ($\mu=E_f$), $\kappa_e$ is less than 10\% of $\kappa_{ph}$. Therefore, $\kappa_e$ is always neglected in the study of thermal conductance of graphene. This situation is applicable to many other 2D materials. However, for stanene (red lines), we find that electron thermal conductance is comparable to phonon thermal conductance at room temperature. When the stanene is gated, $\kappa_e$ can even be larger than $\kappa_{ph}$. Therefore the effect of electron thermal conductance can no longer be neglected.  To have a deeper understanding of the competition between $\kappa_{ph}$ and $\kappa_e$, we analyze their temperature profile as shown in Fig.~\ref{fig:thermal}(b). We find  that $\kappa_{ph}$ increases initially and then saturates with increasing temperature, while $\kappa_e$ increases with increasing temperature following $T^2$ as
\begin{equation}
\sigma_e=\frac{C_2C_0-C_1^2}{C_0}\eta k_B^3T^2.
\end{equation}
 There exists a crossover temperature at $T\approx 342K$ (w/o SOC) and $T\approx 331K$ (w/ SOC). The crossover regime is near the room temperature. which is significantly lower than that of graphene ($\approx 2500K$) \cite{Hossain2015}. According to the saturated phonon thermal conductance of Eq.~(\ref{eq:kph}) and analytical results of $C_n$, we can estimate that the crossover temperature is around
\begin{equation}
T_c=\frac{\hbar}{2k_B}\frac{C_0}{C_2C_0-C_1^2}(\bar{\mathcal{T}}v_F)^{\frac{1}{2}}\left(\frac{K_m}{m}\right)^{\frac{1}{4}}.
\end{equation}
This formula indicates that materials with a strong force constant, light atomic mass and large Fermi velocity will have a large crossover temperature. Using this formula, it is found that the crossover temperature for stanene is $T_c=338K$, which is close to the simulated value of $342K$.

\vspace{10pt}
\label{Conclusion}
\noindent\textit{Conclusion} We have investigated the thermal transport properties of stanene using NEGF approach in combination with first-principles calculations. It is found that stanene possesses a very low phonon thermal conductance in comparison with graphene. Interestingly, the electron thermal conductance of stanene does not follow the conventional Wiedemann-Franz law due to the existence of Dirac point. We have hence derived an analytical formula of the conductance ratio for Dirac fermions from NEGF approach and obtained a modified WF law. The modified WF law predicts a new proportionality constant of $ 5.79\times10^{-8}W\Omega K^{-2}$, which is about two times larger than the Lorenz constant derived from Sommerfeld theory. Importantly, this new constant is applicable to 2D Dirac Fermion systems in general, including graphene. Using this analytical approach, we have also derived an analytical formula to determine the crossover from the phonon dominated regime to the electron dominated regime. Remarkably, this crossover for stanene occurs around room temperature. Hence,  the contribution from the electron thermal conduction in stanene at room temperature is no longer negligible. The fascinating behaviours revealed here in stanene not only widen our knowledge in thermal transport in 2D materials, but also provide a new route to manipulate the thermal conductance of stanene since controlling electrons is much easier than controlling phonons. 

This work was supported in part by a grant from the Science and Engineering Research Council (152-70-00017). The authors gratefully acknowledge the financial support from the Agency for Science, Technology and Research (A*STAR), Singapore and the use of computing resources at the A*STAR Computational Resource Centre, Singapore.

\appendix
\section{Calculation of transmission function} 
The transmission function is an important quantity in the Landauer formalism in the ballistic transport regime. It characterizes the transmission probabilities of phonons coming from left region to right region, scattered by a central region. In this work, all left, central and right regions are monolayer stanene sheet with infinite width. We partition the stanene into periodic blocks in the transport direction for all regions, so that the atoms in each block only interact with those in its two nearest neighbor blocks. Within each block, the lattice has periodic structure in the transverse direction. So the force constants between two atoms can be written as $K_{(l,\vn,b),(l',\vn',b')}$, where $l$, $l'$ are the block indices of each atoms respectively, $\vn$, $\vn'$ are the vectors pointing to the unit cells within each block and $b$, $b'$ are the atomic indices within each unit cell. By making use of the periodicity in the transverse direction, we can transform the force constants into reciprocal lattice in the transverse direction
\begin{equation}
\tk_{(l,b)(l',b'),\vk}=\sum_{\vn}K_{(l,\vn,b)(l',0,b')}e^{-i\vk\cdot\vn}.
\end{equation}
Then the retarded Green's function of the central region can be written as
\begin{equation}
G^r(\omega,\vk)=[(\omega+i\eta)^2M-\tk^C_{\vk}-\Sigma^r_L(\omega,\vk)-\Sigma^r_R(\omega,\vk)]^{-1},
\end{equation}
where $M$ is a diagonal matrix of atomic mass, the superscript $C$ in $\tk^C_{\vk}$ means that the block indices $l$ and $l'$ are inside the central region. $\eta$ is a small positive number and $\Sigma^r_{L/R}$ are the self-energies of the left and right leads. The self-energies can be evaluated from $\Sigma_\alpha^r(\omega,\vk)=V^{\alpha C}_{\vk}g^r_\alpha(\omega,\vk)V^{C\alpha}_{\vk}$ for $\alpha=L,R$, where $V^{\alpha C}$ is the Fourier-transformed coupling matrix between the lead $\alpha$ and central region and $V^{C\alpha}$ is its hermitian conjugate. Here $g^r_\alpha(\omega,\vk)$ is the surface retarded Green's function of lead $\alpha$. It is evaluated according to the generalized eigenvalue problem, as demonstrated in Ref.\cite{Wang2008}. Hence the transmission function can be calculated using
\begin{equation}
\Xi_{ph}[\omega,\vk]=\mbox{Tr}\Big[G^r(\omega,\vk)\Gamma_L(\omega,\vk)G^a(\omega,\vk)\Gamma_R(\omega,\vk)\Big],
\end{equation}
where $G^a(\omega,\vk)=[G^r(\omega,\vk)]^{\dagger}$ is the advanced Green's function and $\Gamma_\alpha(\omega,\vk)=i\Big(\Sigma^r_\alpha(\omega,\vk)-[\Sigma^r_\alpha(\omega,\vk)]^\dagger\Big)$ is the spectral density of the leads.

The energy-dependent transmission function can be evaluated by integrating the mode-dependent transmission function over all the modes. For 2D materials, in the transverse direction, $\vk$ is a one-dimension vector so that it is actually a number, denoted as $k_\parallel$
\begin{equation}
\mathcal{T}_{ph}(\omega)=a\int_{k_{\parallel}\in\mbox{BZ}}\frac{dk_\parallel}{2\pi}\Xi_{ph}(\omega,k_\parallel).
\end{equation}

For the electron transmission function, the basic formalism is similar, except that instead of starting from the force constant matrix, the electron formalism starts from the hopping matrix from the tight-binding model. The frequency dependence of phonon transmission should be changed to energy dependence of electrons by replacing $(\omega+i\eta)^2$ with $E+i\eta$ in the calculation of retarded Green's function. The mass matrix $M$ becomes the overlap matrix denoting the overlapping of electron wavefunctions between the sites. In this calculation we use identity assuming electrons are not overlapped. When spin-orbital coupling is not considered, a factor of 2 is multiplied in order to take care of spin degeneracy.

\bibliography{ST}
\end{document}